\documentclass[prd,aps,10pt,showkeys,nofootinbib,showpacs,twocolumn]{revtex4-2}

\usepackage{amsmath,amssymb,amsfonts,amsthm}
\usepackage{physics}
\usepackage{xcolor}
\usepackage{graphicx}
\usepackage{hyperref}
\usepackage{cleveref}

\usepackage{tikz}
\usetikzlibrary{decorations.pathmorphing,patterns}

\newcommand{\im}{{\rm i}}

\newcommand{\td}{\text{d}}

\newcommand{\sL}{\sigma_{\Lambda}}

\def\be{\begin{equation}}
\def\ee{\end{equation}}

\begin{document}

\title{Cyclic Kruskal Universe: a quantum-corrected Schwarzschild black hole in unitary unimodular gravity}

\author{Steffen Gielen}
\affiliation{School of Mathematical and Physical Sciences, University of Sheffield, Hicks Building, Hounsfield Road, Sheffield S3 7RH, United Kingdom}
\email{s.c.gielen@sheffield.ac.uk} \email{sried1@sheffield.ac.uk}
\author{Sofie Ried}
\affiliation{School of Mathematical and Physical Sciences, University of Sheffield, Hicks Building, Hounsfield Road, Sheffield S3 7RH, United Kingdom}
\date{\today}

\begin{abstract}
We analyse the physical properties of an analytical, nonsingular quantum-corrected black hole solution recently derived in a minisuperspace model for unimodular gravity under the assumption of unitarity in unimodular time. We show that the metric corrections compared to the classical Schwarzschild solutions only depend on a single new parameter, corresponding to a minimal radius where a black hole-white hole transition occurs. While these corrections substantially alter the structure of the spacetime near this minimal radius, they fall off rapidly towards infinity, and we show in various examples how physical properties of the exterior spacetime are very close to those of the Schwarzschild solution. We derive the maximal analytic extension of the initial solution, which corresponds to an infinite sequence of Kruskal spacetimes connected via black-to-white hole transitions, and compare with some other proposals for non-singular black hole metrics. The metric violates the achronal averaged null energy condition, which indicates that we are capturing physics beyond the semiclassical approximation. Finally, we include some thoughts on how to go beyond the simple eternal black hole-white hole model presented here.
\end{abstract}


\maketitle

\section{Introduction}

Black holes have been a particular focus of research in gravitational physics for many years, from both an observational and a theoretical point of view. Recent observations such as those made by the Event Horizon Telescope \cite{EventHorizonTelescope:2022wkp} now allow us to compare the features of black holes seen in our Universe to those predicted by general relativity or modified theories of gravity; on the theoretical side, black holes are often seen as perhaps unique windows into quantum gravity because of their thermodynamic properties \cite{Hawking:1975vcx}, and because they contain a singularity where the laws of physics break down. Insights from quantum gravity are presumably needed to extend this classically singular picture, leading to a range of possible outcomes. A complete understanding of black holes in quantum gravity might also shed light on the famous information loss puzzle of black hole radiation \cite{Hossenfelder:2009xq}, as it should include the final stages of evaporation.

The simplest non-rotating black hole solutions in classical relativity are spherically symmetric and moreover, in the simplest case of vacuum solutions, static by Birkhoff's theorem. This high amount of symmetry allows deriving the exact Schwarzschild solution straightforwardly. Assuming the same symmetries might then also lead to a derivation of quantum-corrected models of black holes in theories of quantum gravity, which can potentially resolve the classical singularity. In the setting of (canonical) loop quantum gravity (LQG), this approach has led to a number of proposals \cite{Gambini:2013hna,Kelly:2020uwj,Ashtekar2023,Alonso-Bardaji:2022ear,Belfaqih:2024vfk,Lewandowski:2022zce,Han:2023wxg} with different physical properties, and to an active debate as to the merits of different models and relation to a possible -- so far incomplete -- theory of LQG. Such models can also be compared with results from other approaches such as asymptotic safety \cite{EichhornHeld} or string theory \cite{Nicolini}.

In our previous work \cite{Gielen:2025ovv} we studied a spherically symmetric and static model of unimodular gravity, quantised in the usual canonical Wheeler--DeWitt formalism. In the Hamiltonian setting, one evolves in a radial coordinate that is spacelike outside, but timelike inside the horizon. Crucially, we then imposed unitary evolution with respect to the unimodular ``time'' or radial coordinate preferred by the theory. Since the classical singularity can be reached in finite time as measured by this clock, unitarity requires a reflecting boundary condition resolving the singularity (see Refs.~\cite{Gielen:2020abd,Gielen:2024lpm} for other examples of this behaviour), and the black hole transitions to a time-reflected asymptotic white hole solution. The result of this quantisation is hence a model that shares some of the features of LQG-inspired black hole models, but derived in a rather different way: the corrections in the metric compared to classical solutions are directly quantum as they arise from a boundary condition associated to unitarity, and do not have a direct interpretation as higher-curvature or discreteness corrections that one could add to the classical theory.

The main purpose of this paper is to illustrate the physical properties of the analytical solution derived in Ref.~\cite{Gielen:2025ovv}, in a particular case where the quantum state is chosen to be very semiclassical and the expectation value of the cosmological constant (arising as an integration constant in unimodular gravity) is zero. We write the metric in more familiar Schwarzschild-like coordinates where the corrections to the classical solution are manifest (Sec.~\ref{sec:coord}), and proceed to take a closer look at global properties (characteristic length scales, curvature invariants, geodesic completeness) in Sec.~\ref{sec:global}. We also consider an analytic extension of the metric, analogous to the classic Kruskal extension of the Schwarzschild solution, and mention the extension of our analysis to a nonvanishing cosmological constant. In Sec.~\ref{sec:energy} we show that our spacetime violates the achronal averaged null energy condition (AANEC) and reflect on what that implies. We conclude in Sec.~\ref{sec:conc} by summarising our results and comparing them to other proposals for quantum-corrected black hole metrics. We will be working in units where $c = \hbar = 1$.

\section{Quantum-corrected Schwarzschild metric}
\label{sec:coord}

In Ref.~\cite{Gielen:2025ovv}, we studied spherically symmetric and static spacetimes starting from the line element 
\begin{align}
\label{equ:lineel}
    \td s^2 = -\frac{\eta(t)N(t)^2}{\xi(t)}\td t^2 + \frac{\xi(t)}{\eta(t)} \td z^2 + \eta(t)^2\td\Omega^2\,.
\end{align}
Here the function $\eta(t)$ is assumed to be non-negative, given that it measures the geometric radius of 2-spheres of constant $(t,z)$ which have area $4\pi\eta^2$. On the other hand, $\xi(t)$ is allowed to be positive or negative, corresponding to both the interior and exterior parts of a black hole solution. When $\xi(t)<0$, one can see that $t$ becomes spacelike and the system now describes ``evolution'' in a radial direction. The lapse $N(t)$ is initially kept general, so that reparametrisation invariance with respect to transformations $t\rightarrow \tilde{t}(t)$ is retained.

The metric is not well-defined at $\xi=0$, which marks the Schwarzschild horizon coordinate singularity. Crucially, however, the minisuperspace variables $\eta$ and $\xi$ remain well-defined across their specified range. Therefore, mathematically, there is no obstruction to a unified quantisation of both the interior and exterior regions, an approach also taken in Ref.~\cite{Cavaglia:1994yc,Vakili:2011nc} (related ideas are also presented in Ref.~\cite{Hartnoll:2022snh}, for example). For more discussion of the appropriate range for $\xi$ and a brief analysis of a quantisation restricted to the interior, see Ref.~\cite{Gielen:2025ovv}. At the level of minisuperspace models, there seems to be a choice between a model that includes both exterior and interior and a model that only describes the interior.

In passing to the Hamiltonian setting, one now assumes that $t$ can be used as an evolution parameter throughout, so that $t$-derivatives appear as velocities in the Legendre transform (this generalisation of Hamiltonian methods to potentially spacelike ``evolution'' also appears, for example, in the context of AdS/CFT \cite{deBoer:2000cz}). One obtains a very simple Hamiltonian
\be
H = -N \left[ \pi_\eta \pi_\xi +1 -\eta^2\Lambda \right]
\ee
where, given that we are in unimodular gravity, $\Lambda$ appears as a physical degree of freedom conjugate to a ``clock'' $T$, interpreted as unimodular time. The clock can be used to construct relational observables such as $\eta(T)$ and $\xi(T)$, i.e., metric coefficients expressed relative to the physical clock $T$, which is a dynamical variable of the system rather than an arbitrary coordinate. This procedure of moving from coordinate-dependent solutions to those expressed relative to the clock $T$ can also be seen as fixing the gauge parameter $N$ via the equation of motion
\be
\dot{T} = \{T,H\} = N\eta^2 \stackrel{!}{=} 1\,.
\label{equ:gaugecond}
\ee

Classical solutions can be derived as
\begin{align}
 \eta(T) & = \sqrt[3]{3kT}\,,\nonumber
 \\ \xi(T) & = \frac{\Lambda}{k}T - \sqrt[3]{\frac{3T}{k^5}} + \xi_0\,,
 \label{eq:classsol}
\end{align}
where $k\neq 0$ and $\xi_0$ are integration constants appearing in addition to the conserved $\Lambda$. The black (white) hole mass associated to the solution is given by $k^2\xi_0=2GM$. If we take $T$ as defining an arrow of time, the solutions with $k>0$ are only defined  for positive $T$, corresponding to a white hole emerging from the singularity, and those with $k<0$ are only defined for negative $T$ corresponding to a black hole meeting the singularity in its future.

In the quantum theory, $\eta(T)$ and $\xi(T)$ are represented as operators. One can find quantum states $\psi_{\Lambda,k}$, which are exact solutions of the Wheeler--DeWitt equation, for each choice of constants $(\Lambda,k)$, so that the general solution is
\be
\psi(\eta,\xi,T) = \int_{-\infty}^\infty \frac{{\rm d}\Lambda\,{\rm d}k}{(2\pi)^2}\,\alpha(\Lambda,k)\psi_{\Lambda,k}(\eta,\xi,T)\,.
\ee
While $\alpha(\Lambda,k)$ is initially completely arbitrary, self-adjointness of the Hamiltonian (and hence unitarity of $T$ evolution) requires choosing a subspace such that
\be
\alpha(\Lambda,-k)= e^{\im \chi(k)}\alpha(\Lambda,k)
\label{equ:alphacondition}
\ee
for some function $\chi(k)$. Since $T$ is not timelike outside the horizon, requiring unitary time evolution is not enough to justify imposing unitarity in $T$. One may note, however, that for positive mass black (white) holes $T$ is timelike close to the singularity, where imposing unitarity is going to have the largest effect. Another argument for imposing unitarity is that the Hamiltonian is minus the cosmological constant operator $\hat{\Lambda}$. Hence imposing unitarity in $T$ is equivalent to making $\hat{\Lambda}$ a self-adjoint operator, and therefore an observable.

Requiring that the allowed quantum solutions have a symmetry associated to time reversal of the classical solutions further constrains the function $\chi$ to $\chi(k)=\beta k$, where $\beta$ is a free parameter of the theory.

For a semiclassical analysis, we chose Gaussian wavepackets defined by 
\begin{align}
\label{equ:alphasc}
    \alpha_{sc}(\Lambda,k) = \mathcal{N} e^{-\im\frac{\beta}{2} k} e^{-\frac{(|k|-k_c)^2}{2\sigma_k^2}}e^{-\frac{(\Lambda-\Lambda_c)^2}{2\sigma_{\Lambda}^2}}\,,
\end{align}
where $\mathcal{N}$ is a normalisation factor and the amplitude function is chosen to satisfy Eq.~(\ref{equ:alphacondition}). Such a state is determined by four free parameters, the expectation values $\Lambda_c\in\mathbb{R}$ and $k_c>0$ and variances given by $\sigma_k^2$ and $\sigma_\Lambda^2$. One can then derive analytical expectation values in any such state \cite{Gielen:2025ovv}, but the resulting expressions are rather complicated. To simplify matters, one can assume that the state is sharply peaked in $k$, so that $\sigma_k\ll k_c$, as seems reasonable for a semiclassical state. In this limit, one can compute approximate expectation values 
\begin{align}
\langle\hat\eta(T)\rangle & \approx \frac{1}{\sqrt{\pi}}\left(\frac{3 k_c}{\sigma_\Lambda}\right)^{1/3}\Gamma\left(\frac{2}{3}\right){}_1F_1\left(-\frac{1}{6},\frac{1}{2},-\sL^2T^2\right)
\,,\nonumber
\\ \langle\hat\xi(T)\rangle & \approx \frac{\Lambda_c}{\sqrt{\pi}\sigma_\Lambda k_c}e^{-\sigma_\Lambda^2 T^2} + \frac{\Lambda_c}{k_c} T\,{\rm erf}(\sigma_\Lambda T)\nonumber
\\ & \quad +\frac{\Gamma\left(-\frac{1}{3}\right){}_1F_1\left(-\frac{1}{6},\frac{1}{2},-\sL^2T^2\right)}{\sqrt{\pi}\sqrt[3]{9 k_c^5 \sigma_\Lambda}} + \frac{\beta}{2}
\,,
\label{eq:etaxiexpv}
\end{align}
where ${}_1 F_1$ are hypergeometric functions and ${\rm erf}$ is the error function.

These expectation values can then be used to define a quantum-corrected metric via
\begin{align}
    \td s^2 = -\frac{\td T^2}{\langle\hat\eta(T)\rangle^3\langle\hat\xi(T)\rangle} + \frac{\langle\hat\xi(T)\rangle}{\langle\hat\eta(T)\rangle} \td z^2 + \langle\hat{\eta}(T)\rangle^2\td\Omega^2
    \label{equ:quantumss}
\end{align}
using the gauge condition (\ref{equ:gaugecond}) so that $N=1/\eta^2$. Eq.~(\ref{equ:quantumss}) with Eq.~(\ref{eq:etaxiexpv}) can be seen as a definition of a quantum Schwarzschild--(Anti-)de Sitter metric in unitary unimodular gravity under the assumptions we have made; in particular, we quantised only the variables of a symmetry-reduced model, and assumed the gauge condition (\ref{equ:gaugecond}) without corrections at the quantum level. We have also assumed that the state is semiclassical enough to justify the replacement of, e.g., the expectation value of a quantum operator $\widehat{(\xi/\eta)}$ by the ratio of elementary expectation values derived in Eq.~(\ref{eq:etaxiexpv}) (see Ref.~\cite{Gielen:2025ovv} for more detailed discussion).

In Ref.~\cite{Gielen:2025ovv} it was shown that the asymptotic forms of Eq.~(\ref{eq:etaxiexpv}) (as $|T|\rightarrow\infty$) are
\begin{align}
    \langle\hat\eta(T)\rangle & \sim \sqrt[3]{3k_c|T|}\,,\nonumber
   \\  \langle\hat\xi(T)\rangle & \sim \frac{\Lambda_c}{k_c}|T|  - \sqrt[3]{\frac{3|T|}{k_c^5}} + \frac{\beta}{2}\,.
   \label{eq:asymptotics}
\end{align}
These correspond exactly to the classical solutions given in Eq.~(\ref{eq:classsol}), more precisely a black-hole solution with $k<0$ for large negative $T$ and a white-hole solution with $k>0$ for large positive $T$. We also need to identify the self-adjoint extension parameter $\beta$ with the classical integration constant $2\xi_0$. Since the solution stays regular for all $T$, it describes a black-hole to white-hole transition, for which the classical Schwarzschild--(Anti-)de Sitter metric is recovered asymptotically.

In the following, we will primarily be interested in the case $\Lambda_c=0$, for which the state is peaked around  a vanishing cosmological constant. In this case the first two terms in $\langle\hat{\xi}(T)\rangle$ disappear and our metric represents a quantum-corrected Schwarzschild solution, which we will now study as an effectively classical spacetime metric.

One might ask whether, in a limit in which the black (white) hole mass goes to zero, we recover flat Minkowski spacetime without quantum corrections. Classically the mass depends on the product $k^2\xi_0$, but since our solutions are only defined for $k\neq 0$, the limit to take is $\xi_0\rightarrow 0$. (The case $k=0$ is only defined for $\Lambda>0$, where it corresponds to the Nariai spacetime \cite{Gielen:2025ovv}). In the quantum theory, we must similarly assume that $k_c>0$, but here the role of the classical integration constant $\xi_0$ is played by the self-adjoint extension parameter $\beta$. Taking $\beta\rightarrow 0$ leads to a theory in which there are no semiclassical states with nonzero mass, but that theory does feature ``quantum-corrected'' Minkowski solutions in which the coordinate singularity at $T=0$ (corresponding to the origin in polar coordinates) is resolved: there is now a finite minimal radius at which Minkowski spacetime is cut off and glued onto a second copy of Minkowski spacetime. This is because imposing unitarity resolves all singularities that the clock reaches in finite time, including coordinate singularities (see also Refs.~\cite{Gielen:2022tzi,Gielen:2024qml} for  discussion of similar phenomena, and Ref.~\cite{Gielen:2025ovv} for discussion of the same feature for de Sitter spacetime). This type of quantum correction might seem unphysical and can be excluded by taking $\beta\neq 0$. In that case, one can take a different limit (see below) at the level of the effectively classical spacetime metric (\ref{equ:quantumss}).

\section{Global properties of quantum-corrected Schwarzschild}
\label{sec:global}

From now on, we will treat the quantum Schwarzschild metric (\ref{equ:quantumss}) with Eq.~(\ref{eq:etaxiexpv}) (and $\Lambda_c=0$) as an effectively classical spacetime and analyse its properties in detail.

\subsection{Minimal radius and horizon structure}
\label{sec:length}

Eq.~(\ref{equ:quantumss}) is not written in particularly convenient coordinates. If we introduce a radial Schwarzschild coordinate
\be
r(T) = \langle\hat\eta(T)\rangle\,,
\ee
we observe that
\be \xi(r)=\langle\hat\xi(T)\rangle|_{T\rightarrow T(r)} = -\frac{r}{k_c^2}+\frac{\beta}{2} \,.
\ee
If we also rescale $z\rightarrow k_c z$, the metric takes the new form
\be
{\rm d}s^2 = - \left(1-\frac{r_H}{r}\right){\rm d}z^2 +\frac{{\rm d}r^2k_c^2\,T'(r)^2}{r^4\left(1-\frac{r_H}{r}\right)} + r^2\,{\rm d}\Omega^2
\label{newmetric}
\ee
with $r_H:=\frac{1}{2}\beta k_c^2$. $T(r)$ now denotes the implicitly defined inverse function of our definition for $r(T)$, and one can see that all the nontrivial behaviour in the metric is now hidden in this implicit definition and in the global properties of the map from $T$ to $r$. In particular, notice that $g_{zz}$ has the same form as in the classical Schwarzschild solution, as $T'(r)^2$ only appears in the $g_{rr}$ component.

$r(T)$ has a global minimum at $T=0$,
\be
r_{{\rm min}}=r(0)= \frac{\Gamma\left(\frac{2}{3}\right)}{\sqrt{\pi}}\left(\frac{3k_c}{\sigma_\Lambda}\right)^{1/3}\,,
\ee
and is monotonically increasing $(T>0)$ or monotonically decreasing $(T<0)$ away from the minimum. The solution has a clear interpretation as a classical black hole solution contracting to a minimal radius and then ``bouncing'' to an asymptotic white hole solution, as detailed in Ref.~\cite{Gielen:2025ovv}. For our present purposes, this means that the coordinate $r$ is only defined for $r\ge r_{{\rm min}}$ and it can only cover one half of the solution, either corresponding to the white-hole or to the black-hole part. This is as expected from a Schwarzschild-like radial coordinate in such a time-symmetric bounce scenario. Since $r'(0)=0$, we have $|T'(r)|\rightarrow\infty$ as $r\rightarrow r_{{\rm min}}$ and the metric becomes singular there. This is a coordinate singularity and not a curvature singularity, since everything is regular in the original coordinate system. Indeed the spacetime is free from any curvature singularities, as we will confirm explicitly below. Away from $r=r_{{\rm min}}$, $T'(r)$ is always positive or always negative and finite.

Hence, the only point at which the metric is ill-defined is $r=r_H$, which we identify with an horizon analogous to the usual Schwarzschild horizon. This horizon is only part of the spacetime if $r_H>r_{{\rm min}}$, and it only applies to the part of the spacetime covered by the new coordinate patch, which is only one half of the total solution. In this model $r_{{\rm min}}$ is a free parameter determined by properties of the state, and unrelated to $r_H$ or to any fundamental scale such as the Planck scale.

To see explicitly that $r=r_H$ denotes an horizon, one can introduce ingoing Eddington--Finkelstein coordinates in the usual way via
\be
\label{equ:EFtrafo}
{\rm d}z = {\rm d}v - {\rm d}r\,\frac{|T'(r)|\,k_c}{r^2\left(1-\frac{r_H}{r}\right)}\,,
\ee
so that
\be
{\rm d}s^2 = - \left(1-\frac{r_H}{r}\right){\rm d}v^2 + \frac{2 |T'(r)|k_c}{r^2}\,{\rm d}v\,{\rm d}r  + r^2\,{\rm d}\Omega^2\,.
\label{EF}
\ee

In these coordinates we have removed the coordinate singularity at $r=r_H$. Then for ingoing future-directed null or timelike curves we have
\begin{align}
U^2 & = -\left(1-\frac{r_H}{r}\right)\left(\frac{{\rm d}v}{{\rm d}\lambda}\right)^2 + \frac{2 |T'(r)|k_c}{r^2}\,\left(\frac{{\rm d}v}{{\rm d}\lambda}\right)\left(\frac{{\rm d}r}{{\rm d}\lambda}\right)\nonumber
\\ & \quad + r^2  \left(\frac{{\rm d}\Omega}{{\rm d}\lambda}\right)^2 \le 0\,,
\end{align}
where $U^\mu = {\rm d}x^\mu/{\rm d}\lambda$ is the tangent vector to the curve, and hence for $r<r_H$ the second term must be negative; curves satisfying $v'(\lambda)\ge 0$ must then have $r'(\lambda)\le 0$ and hence $r$ can only decrease towards $r=r_{{\rm min}}$. Thus $r=r_H$ still appears to be a horizon as far as this coordinate patch is concerned. This argument is independent of the form of the function $|T'(r)|$ and identical to the usual argument showing the presence of a horizon in Schwarzschild \cite{HawkingEllis}.

In analogy with the Schwarzschild event horizon, we can associate a temperature to the black hole by the usual calculation extending the exterior spacetime into Euclidean signature. The Euclidean metric ($r>r_H$) is
\be
{\rm d}s^2 = \left(1-\frac{r_H}{r}\right){\rm d}\tau^2 +\frac{T'(r)^2\,k_c^2}{r^3\left(r-r_H\right)}\,{\rm d}r^2 + r^2\,{\rm d}\Omega^2\,.
\ee
Near $r=r_H$, we can write $r=r_H+\alpha R^2$ and obtain, approximately for $\alpha R^2 \ll r_H$,
\be
{\rm d}s^2 = \frac{\alpha R^2}{r_H}\, {\rm d}\tau^2 + \frac{4\alpha\,T'(r_H)^2\,k_c^2}{r_H^3}\,{\rm d}R^2 + r_H^2\,{\rm d}\Omega^2\,.
\ee
Then by setting $\alpha=r_H^3/(4 T'(r_H)^2 k_c^2)$, this metric takes the form
\be
{\rm d}s^2 = \frac{r_H^2 R^2}{4 T'(r_H)^2 k_c^2}\, {\rm d}\tau^2 + {\rm d}R^2 + r_H^2\,{\rm d}\Omega^2\,.
\ee
This metric has a conical singularity at $R=0$ unless $\tau$ is an angular coordinate with periodicity
\be
\label{eq:htemp}
\beta_{{\rm per}} = \frac{4\pi k_c |T'(r_H)|}{r_H}
\ee
which defines the inverse temperature. It explicitly depends on $T'(r)$, which we will look at more closely now.

\subsection{Properties of $T'(r)$}
\label{sec:T}

While $T'(r)$ is only known as an implicit function, one can make several statements about its properties and hence the properties of the metric component $g_{rr}$, which now captures all the quantum corrections.

From the asymptotic form (\ref{eq:asymptotics}) valid for large $|T|$, we find that at large $r$
\be
|T(r)|\sim \frac{r^3}{3k_c}\,,\quad |T'(r)|\sim \frac{r^2}{k_c}\,,
\ee
which reduces Eq.~(\ref{newmetric}) to the classical Schwarzschild metric and Eq.~(\ref{eq:htemp}) to the familiar expression $\beta=4\pi r_H$. Away from this asymptotic limit, there are corrections, but all corrections depend only on $r_{{\rm min}}$ via the ratio $\frac{r}{r_{{\rm min}}}$ and not independently on the state parameters $(k_c,\sigma_{\Lambda})$ as it might seem at first.

To see this, we can write $r(T)=r_{{\rm min}} F(\sigma_\Lambda^2T^2)$ where $F$ is a shorthand for the hypergeometric function in Eq.~(\ref{eq:etaxiexpv}). Then it follows that 
\be
T^2 = \frac{1}{\sigma_\Lambda^2}F^{-1}\left(\frac{r}{r_{{\rm min}}}\right) = \left(\frac{\sqrt{\pi}}{\Gamma(\frac{2}{3})}\right)^6 \frac{r_{{\rm min}}^6}{9k_c^2}F^{-1}\left(\frac{r}{r_{{\rm min}}}\right)
\ee
where $F^{-1}$ is the inverse function of $F$. Hence we see that $T=\frac{r_{{\rm min}}^3}{k_c}\cdot G(\frac{r}{r_{{\rm min}}})=\frac{r^3}{k_c}\cdot H(\frac{r}{r_{{\rm min}}})$ for some function $G$ and some other function $H$.

This means we can write the metric entirely using $r_{{\rm min}}$ and $r_H$ instead of using $k_c$, $\sigma_{\Lambda}$ and $\beta$, meaning these three free parameters include only two physically relevant variables. By setting $|T'(r)|=\frac{r^2}{k_c}\tau(\frac{r}{r_{{\rm min}}})$, we could write
\begin{align}
    \td s^2 = -\left(1-\frac{r_H}{r}\right)\td z^2+\frac{\tau\left(\frac{r}{r_{{\rm min}}}\right)^2}{\left(1-\frac{r_H}{r}\right)}\td r^2+r^2\td\Omega^2\,,
\end{align}
where the function $\tau$ is still only known implicitly, but can be expanded up to any desired order in a perturbative expansion. 
As a consequence, we can see that the self-adjoint extension parameter $\beta$ (for $\beta>0$) does not really play a distinguished role in giving inequivalent quantum corrections, as one might have expected. A state specified by $(k_c,\sigma_{\Lambda})$ in the theory with $\beta$ and a state defined by $(\frac{1}{\mu}k_c,\frac{1}{\mu}\sigma_{\Lambda})$ in the theory with $\tilde{\beta} = \mu^2\beta$ give the same expressions for $r_{{\rm min}}$ and $r_H$, thus will give the same quantum-corrected Schwarzschild metric.

While $r_H$ and $r_{{\rm min}}$ are generally just two independent parameters associated to our metric, one could assume that $\beta$ is somehow fixed (it can be seen as a choice of quantum theory, rather than a choice of state) while the state parameters $k_c$ and $\sigma_\Lambda$ are variable. Then, re-expressing $k_c$ in terms of $r_H$ and writing $r_H=2 GM$ in terms of a mass parameter  $M$, we have
\be
\label{eq:rminM}
r_{{\rm min}} \sim \left(\frac{k_c}{\sigma_\Lambda}\right)^{1/3} \sim \frac{(GM)^{1/6}}{\beta^{1/6}\sigma_\Lambda^{1/3}}\,.
\ee
Then, assuming that $\beta$ and $\sigma_\Lambda$ are fixed while we vary $M$, the minimal radius only grows very weakly with $M$. For smaller and smaller black holes, eventually we would have $r_{{\rm min}}=r_H$ and the horizon would no longer be part of the resulting spacetime, as the solution ``bounces'' even before the horizon is reached. We will return to this feature in the discussion below.

From an observational point of view, $r_{{\rm min}}$ should be smaller than the horizon radius of the lightest black hole observed so far, which has about 3.8 solar masses:
\be
r_{{\rm min}} < 11 \;{\rm km}\,.
\ee

Including the leading-order corrections in $\frac{r_{{\rm min}}}{r}$, we have
\begin{align}
\label{equ:Tlarger}
\begin{split}
|T(r)| &\sim \frac{r^3}{3k_c}\left(1+\frac{\pi^3}{6\Gamma\left(\frac{2}{3}\right)^6}\frac{r_{{\rm min}}^6}{r^6}+\frac{\pi^6}{18\Gamma\left(\frac{2}{3}\right)^{12}}\frac{r_{{\rm min}}^{12}}{r^{12}}\right)\,, \\
\tau\left(\frac{r}{r_{\rm min}}\right)& \sim 1-\frac{\pi^3}{6\Gamma\left(\frac{2}{3}\right)^6}\frac{r_{{\rm min}}^6}{r^6}-\frac{\pi^6}{6\Gamma\left(\frac{2}{3}\right)^{12}}\frac{r_{{\rm min}}^{12}}{r^{12}}\,,
\end{split}
\end{align}
where we are neglecting exponentially suppressed terms of $O(e^{-\alpha \frac{r^6}{r_{{\rm min}}^6}}\cdot \frac{r_{{\rm min}}^5}{r^5})$ for some constant $\alpha$. One can see that the corrections are very small for any $r\gg r_{{\rm min}}$, so for models in which $r_{{\rm min}}$ is well inside the horizon one would expect no observable deviations from the classical Schwarzschild solution in the exterior. 

As we approach the minimal radius, $r\rightarrow r_{{\rm min}}$,
\be
\label{equ:Trsmall}
|T(r)|\sim \frac{\sqrt{3}}{\sigma_\Lambda}\sqrt{\frac{r-r_{{\rm min}}}{r_{{\rm min}}}}\,, \quad |T'(r)|\sim \frac{\sqrt{3}}{2\sigma_\Lambda\sqrt{r_{{\rm min}}(r-r_{{\rm min}})}}\,.
\ee
As a confirmation of our previous argument, we can see that $k_c T$ is only a function of $r_{{\rm min}}$ and $r$ also in this limit. These expressions show that the bounce surface at $r=r_{{\rm min}}$ can be reached in finite proper time, since
\be
\int_{r_{{\rm min}}}^{r_H} {\rm d}r\;\frac{|T'(r)|\,k_c}{r^2\sqrt{\frac{r_H}{r}-1}}
\ee
is an integral over a finite interval whose integrand only has singularities of the type $(x-x_0)^{-1/2}$ at the boundary values of the integral, which lead to a finite integral.

As a final remark, we point out that in the case where the mass is taken to be zero, $r_H=0$, we can also take the minimal radius $r_{{\rm min}}$ to zero. Given the definitions of both $r_H$ and $r_{{\rm min}}$, this is a formal limit $k_c\rightarrow 0$, which is well-defined at the level of the metric seen as a correction of the classical Schwarzschild solution. Evidently, such a limit reduces the metric to the classical Minkowski metric without any corrections. Given our remarks above, however, such a solution would not arise from a quantum theory with unitarity in unimodular time, where the limit $k_c\rightarrow 0$ is not defined.

\subsection{Curvature invariants}
\label{sec:kretsch}

To confirm that the Schwarzschild singularity is resolved in our quantum-corrected spacetime, we can calculate the Kretschmann scalar $K=R_{\mu\nu\rho\sigma}R^{\nu\mu\rho\sigma}$, finding
\begin{align}
\begin{split}
    K = &\,\frac{4}{r^4}+\frac{4r^2}{k_c^4T'(r)^4}(9r^2-14r r_H+6 r_H^2)\\
    &+\frac{8}{k_c^2rT'(r)^2}(r_H-r)\\
    &-\frac{8r^3T''(r)}{k_c^4T'(r)^5}(4r-3r_H)(r-r_H)\\
    &+\frac{r^4T''(r)^2}{k_c^4T'(r)^6}(8r^2-16 r r_H+9 r_H^2)\,.
\end{split}
\end{align}
We can see that at $r=r_{{\rm min}}$, where the metric becomes singular in these coordinates, we have $K=4/r_{{\rm min}}^4$, which is finite and thus confirms that the singularity is a mere coordinate artifact. For large $r$ we can expand the Kretschmann scalar as 
\be
    K \sim  \frac{12 r_H^2}{r^6}\left(1+\frac{\pi^3}{3\Gamma\left(\frac{2}{3}\right)^6}\frac{r_{{\rm min}}^6}{r^6}\left(5-\frac{8r}{3r_H}\right)\right)\,.  
\ee
For $r_{\rm min} = r_H$ the correction term at the horizon evaluates to $\frac{7\pi^3}{9\Gamma\left(\frac{2}{3}\right)^6}\approx 3.9$, which is an order one correction. However, the maximal value of this correction terms falls off rapidly when we choose  $r_{{\rm min}}<r_H$. For example if one takes $r_{{\rm min}} = r_H/10$ the correction becomes $O(10^{-6})$. If one assumes that $r_{\rm min}$ is microscopic, e.g., of the order of the Planck scale, all corrections would of course again be very heavily suppressed for macroscopic black holes.

The Ricci scalar of our quantum-corrected Schwarzschild metric is given by
\be
 R = \frac{2}{r^2} + \frac{2r(3r_H-5r)}{k_c^2 T'(r)^2} + \frac{r^2(4r-3r_H)T''(r)}{k_c^2 T'(r)^3}\,.
\ee
Unlike in the classical case, this is non-zero, though it does again fall off very rapidly at large radius,
\be
 R \sim \frac{\pi^3}{3\Gamma\left(\frac{2}{3}\right)^6}\,\frac{r_{{\rm min}}^6}{r^8}\left(10-9\frac{r_H}{r}\right)\,.
\ee
As for the Kretschmann scalar, we can see that there are no values for $r$ for which $R$ diverges.

\subsection{Geodesics}
\label{sec:geo}

Our metric describes a black hole spacetime that smoothly transitions to a white hole at a minimal radius, allowing for travel from an exterior region through the black hole interior, into the white hole interior, and out to another exterior region. We already know that our $r$ coordinate patch only covers one half of the entire spacetime, and geodesics can reach the boundary of this patch at $r=r_{{\rm min}}$ in finite affine parameter. A critical question, however, is whether this geodesic incompleteness also extends to the exterior patch $r>r_H$, or whether that patch covers the exterior all the way to timelike or null infinity. The first scenario would suggest the possibility of another connection between the asymptotic regions associated to the black-hole and white-hole parts of the spacetime. 

We should expect that the coordinate patch $r>r_H$ covers the entire exterior to timelike or null infinity, and that no such connection between asymptotic exterior regions exists. This is indeed what happens in LQG-inspired scenarios such as \cite{Belfaqih:2024vfk,Ashtekar:2005cj} when an eternal black hole is considered. Scenarios involving the ``tunnelling'' of black holes to white holes, as discussed for instance in \cite{Han:2023wxg}, rely on the inclusion of Hawking evaporation leading to a finite rather than infinite lifetime of a black hole (see also our discussion in the conclusions). In our model, there is no Hawking radiation and the metric is identical to Schwarzschild up to corrections that can be very small in the exterior, as discussed above. In the classification of Ref.~\cite{Hossenfelder:2009xq}, our spacetime aligns with ``Option 4: There is an horizon but no singularity''. (Strictly speaking, their discussion includes a star collapsing into a black hole which then evaporates, whereas we are considering an eternal black hole.) Such a configuration includes a ``baby universe'' accessible only through the black hole interior (see, e.g., Ref.~\cite{Frolov:1988vj} for an old proposal of such a baby universe). In our case, the ``original'' and ``baby'' universes are isometric mirror images of each other.

We want to follow the geodesic of a massive (massless) particle to timelike (null) infinity. Timelike ($\kappa=1$) or null ($\kappa=0$) geodesics can be assumed to satisfy
\begin{align}
\label{equ:timelikeu}
    -\kappa &=  - \left(1-\frac{r_H}{r}\right)(U^z)^2 +\frac{k_c^2T'(r)^2}{r^4\left(1-\frac{r_H}{r}\right)}(U^r)^2\nonumber\\ & \quad +r^2((U^{\theta})^2+\sin^2 \theta \,(U^\varphi)^2)\,,
\end{align}
where as before $U^\mu=\td x^\mu / \td \lambda$, and $\lambda$ is proper time in the timelike case or an affine parameter in the null case. We can use the symmetries of our spacetime to simplify this expression. Our metric is static and spherically symmetric, thus $\partial_zg_{\mu\nu} = \partial_{\varphi}g_{\mu\nu} = 0$. The symmetries correspond to Killing vectors and conserved quantities related to the energy and angular momentum given by
\begin{align}
    E = \left(1-\frac{r_H}{r}\right)U^z\,,\quad L = r^2\sin^2\theta\,U^{\varphi}\,.
\end{align}
In addition, we assume initial conditions $U^{\theta}_0 = 0$ and $\theta_0 = \frac{\pi}{2}$ to simplify our expression. The geodesic equations imply that these initial conditions lead to $U^{\theta} = 0$ along the entire geodesic. Eq.~(\ref{equ:timelikeu}) then becomes
\begin{align}
\label{equ:simpu}
   (U^r)^2 = \frac{r^4\left(1-\frac{r_H}{r}\right)}{k_c^2T'(r)^2}\left(\frac{E^2}{1-\frac{r_H}{r}}-\frac{L^2}{r^2}-\kappa\right)\,.
\end{align}
Assuming that our geodesic is moving away from the horizon, $U^r>0$, and integrating leads to the expression
\begin{align}
    \int_0^{\lambda_f}\td\lambda = \int_{r_i}^{r_f}\td r\frac{k_c|T'(r)|}{r^2\sqrt{1-\frac{r_H}{r}}}\left(\frac{E^2}{1-\frac{r_H}{r}}-\frac{L^2}{r^2}-\kappa\right)^{-\frac{1}{2}}\,.
\end{align}
The integrand on the right-hand side is positive everywhere, assuming $r_i>r_H$, and asymptotically reaches the constant $\frac{1}{\sqrt{E^2-\kappa}}$ for $r\rightarrow\infty$. Therefore, the right-hand side diverges when sending $r_f\rightarrow\infty$ and it takes infinite proper time or affine parameter to reach timelike or null infinity. Again, we see that the calculation is very similar to the one for the uncorrected Schwarzschild metric, and the precise form of the function $|T'(r)|$ is not important.

One can also consider bounded geodesics, e.g., circular orbits. In order for an orbit to be circular, we require $U^r = 0$ and $\td U^r/\td \lambda = \td/\td r\left(\frac{1}{2}(U^r)^2\right) = 0$. Thus we impose that both Eq.~(\ref{equ:simpu}) and the $r$ derivative of Eq.~(\ref{equ:simpu}) vanish. If we assume $r_H>r_{{\rm min}}$, so that $T'(r)^2<\infty$ for all $r>r_H$, we end up with the conditions
\begin{align}
    0 &= \frac{L^2}{r^2}-\frac{E^2}{1-\frac{r_H}{r}}+\kappa\,,\\
    0 &= -\frac{2L^2}{r^3}+\frac{E^2r_H}{(r-r_H)^2}\,,
\end{align}
which are identical to the classical conditions for a circular orbit. Again, notice that the specific form of $T'(r)$ does not appear. The solutions are hence the radii at which there are circular geodesic orbits classically:
\begin{align}
    r_O^{\pm} = \frac{L^2}{r_H\kappa}\left(1\pm\sqrt{1-\kappa\frac{3r_H^2}{L^2}}\right)\,.
\end{align}

\subsection{Kruskal extension}
\label{sec:ext}

We have so far seen that the quantum-corrected Schwarzschild metric in the $(r,z)$-coordinates retains a structure similar to that of the classical Schwarzschild metric. Notably, the specific form of the function $T(r)$ was not required in the geodesic calculations of the previous section. This structural similarity raises a natural question: Can we construct an extension of our metric, analogous to the Kruskal extension of the classical Schwarzschild spacetime?

The first step is defining the (modified) tortoise coordinate as we already did when looking at the Eddington--Finkelstein coordinates in Eq.~(\ref{equ:EFtrafo}). We can define
\begin{align}
    r^*(r) = \int_{r_{{\rm min}}}^r \td \tilde{r}\;\frac{|T'(\tilde{r})|k_c}{\tilde{r}^2\left( 1-\frac{r_H}{\tilde{r}} \right)}\,,
\end{align}
where the integral cannot be performed explicitly since we do not have an explicit expression for $T'(r)$. Ingoing and outgoing Eddington--Finkelstein coordinates are then given by
\begin{align}
    v = z+r^*\,,\quad u=z-r^*\,,
\end{align}
and the metric has the form
\begin{align}
    \td s^2 = -\left(1-\frac{r_H}{r}\right)\td v\,\td u + r^2\td\Omega^2\,,
\end{align}
where $r=r(r^*)$ is now defined implicitly.
Exponentiating the Eddington--Finkelstein coordinates, we define
\begin{align}
    U = -e^{-\frac{u}{2r_H}}\,,\quad V=e^{\frac{v}{2r_H}}\,.
\end{align}
This transforms the metric into the form
\begin{align}
    \td s^2 = -4r_H^2\left(1-\frac{r_H}{r}\right)e^{-\frac{r^*}{r_H}}\td U\,\td V+r^2\td\Omega^2\,.
\end{align}
In the final step we define the Kruskal--Szekeres coordinates
\begin{align}
    X = \frac{1}{2}(V+U)\,,\quad T=\frac{1}{2}(V-U)\,,
\end{align}
so that we finally arrive at the metric
\begin{align}
\label{eq:kruskalm}
    \td s^2 = 4r_H^2\left(\frac{r_H}{r}-1\right)e^{-\frac{r^*}{r_H}}(\td X^2-\td T^2)+r^2\td\Omega^2\,.
\end{align}
All the steps after defining the tortoise coordinates are the usual steps performed in Schwarzschild spacetime to arrive at the full Kruskal extension. In the black hole interior the new coordinates are restricted to $0< T^2-X^2\leq 1$ with $T>0$. At the minimal radius we have $T^2-X^2=1$ and $r^*(r_{\rm min})=0$, so the metric is well-defined. It is more tricky to check whether the metric is regular at the horizon.

\begin{figure}
    \centering
    \includegraphics[width=0.8\linewidth]{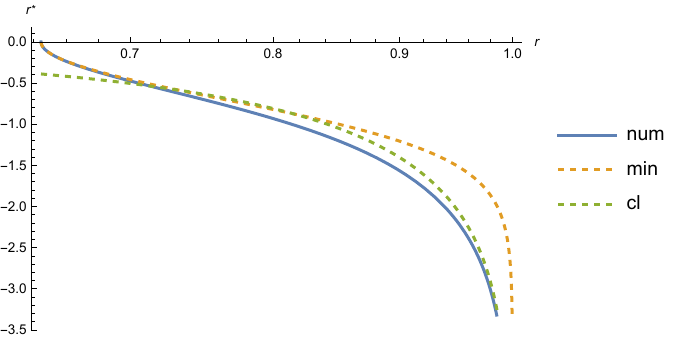}
    \caption{The tortoise coordinate $r^*$ as a function of $r$. The blue solid curve gives the numerical result for the parameters $\beta = 2$, $k_c=1$, $\sigma_{\Lambda} = 5$. The yellow dashed line shows the analytical approximation for $r\approx r_{\rm min}$ and the green dashed line shows the classical solution for $r^*$.}
    \label{fig:tortoise}
\end{figure}

\begin{figure}
    \centering
    \includegraphics[width=0.8\linewidth]{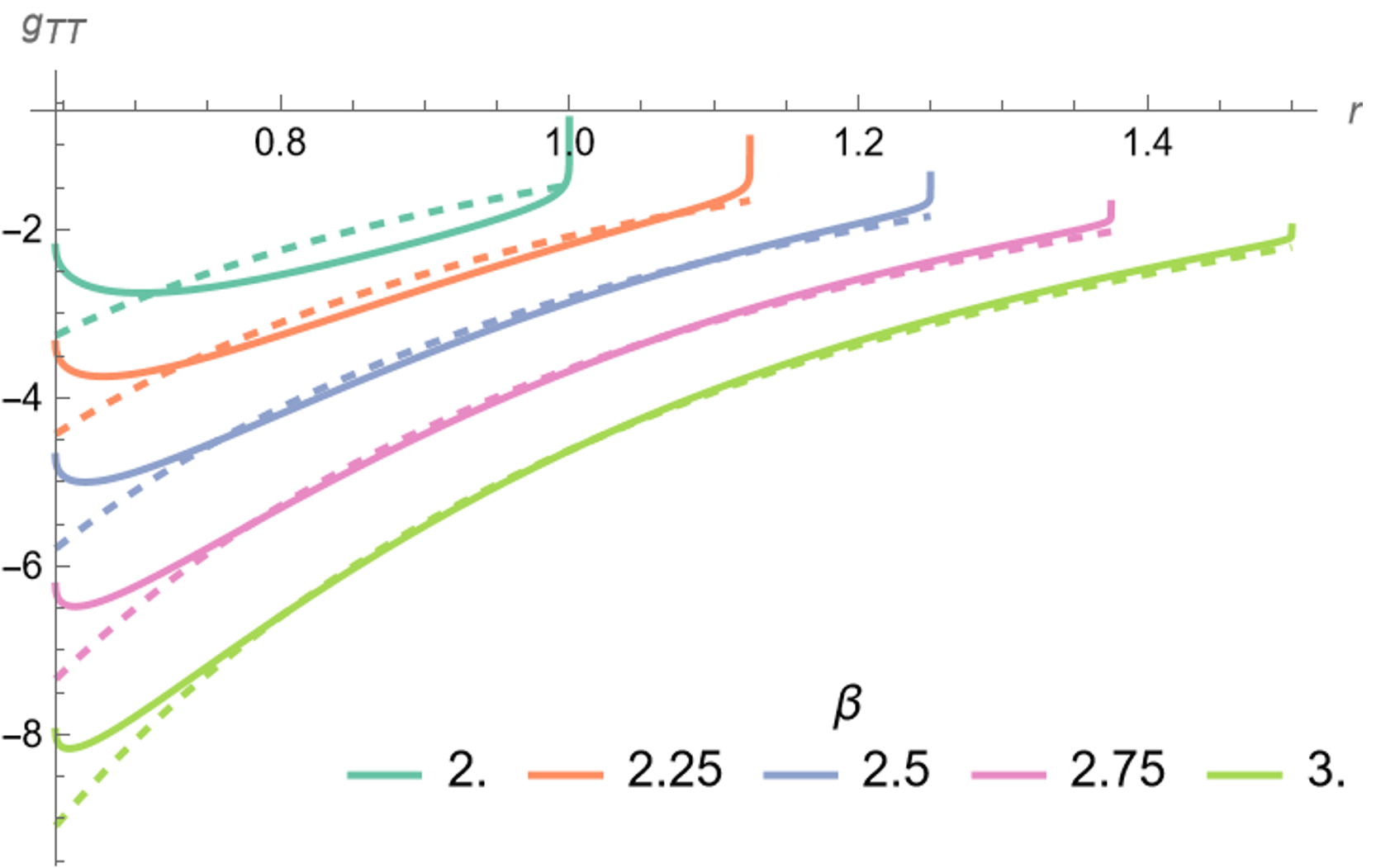}
    \caption{$g_{TT}$ for $r_{\rm min}<r<r_H$. The solid lines show the numerical result for the parameters $2\leq\beta\leq3$, $k_c=1$, $\sigma_{\Lambda} = 5$. The dashed lines show the classical Schwarzschild solution.}
    \label{fig:gTT}
\end{figure}

To make some progress, let us assume that we are interested in $r$ very close to $r_{{\rm min}}$, so that we can use the approximation in Eq.~(\ref{equ:Trsmall}). This leads to the following expression for $r^*$:
\begin{align}
\begin{split}
    r^* \approx& -\frac{\sqrt{3}k_c}{\sigma_{\Lambda}r_Hr_{\rm min}}\left[\arctan\left(\sqrt{\frac{r}{r_{\rm min}}-1}\right)\right.\\
    &\left.+\sqrt{\frac{r_{{\rm min}}}{r_H-r_{{\rm min}}}}{\rm artanh}\left(\sqrt{\frac{r-r_{\rm min}}{r_{H}-r_{\rm min}}}\right)\right]\,.
\end{split}
\end{align}
In contrast the tortoise coordinate for Schwarzschild spacetime is
\begin{align}
    r^* = r+r_H\log\left|\frac{r}{r_H}-1\right|\,.
\end{align}
We can also calculate the tortoise coordinate numerically. A comparison between the three computations can be seen in Fig.~\ref{fig:tortoise}. One can see that the numerical solution agrees with the analytical approximation close to the minimum radius but is much closer to the classical solution at the horizon. This indicates that the behaviour at the horizon of the metric Eq.~(\ref{eq:kruskalm}) is regular. To further confirm this regularity, we numerically plot $g_{TT}$, as shown in Fig.~\ref{fig:gTT}. The results support the conclusion that the metric is indeed well-defined at the horizon. Consequently, it appears that our metric can be extended in the usual way to the coordinate ranges $-\infty < X < \infty$ and $-\infty < T^2 - X^2 \leq 1$. While these observations are not a rigorous proof, being based solely on numerical evidence, they provide strong motivation for such an extension.


Since we have additionally seen that the metric is regular at $X^2-T^2 = 1$, i.e., at $r_{\rm min}$, we can glue together two Kruskal extended spacetimes at $r_{\rm min}$. The original $(T,z)$-coordinates (see Eq.~(\ref{equ:quantumss})) cover both the interior of the black hole and the white hole that was glued on, which ensures that we can make the gluing smooth. In this manner we can glue together infinitely many Kruskal extensions of modified Schwarzschild spacetimes. In this picture, we have an infinite cycle of identical extended black hole spacetimes, which is why we refer to our construction as the \textit{Cyclic Kruskal Universe}.

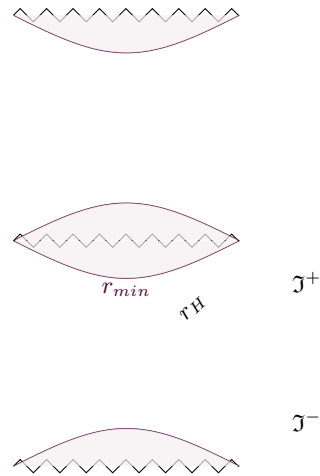
\begin{figure}
\centering
\begin{tikzpicture}[scale=0.5]
    \colorlet{mydarkpurple}{blue!40!red!50!black}
    \colorlet{mylightpurple}{mydarkpurple!80!red!6}

    %defining coordinates
    \coordinate (Il) at (0,-3);
    \coordinate (Ib) at (3,-6);
    \coordinate (Ir) at (6,-3);
    \coordinate (It) at (3,0);
    \coordinate (IIt) at (-3,0);
    \coordinate (IIl) at (-6,-3);
    \coordinate (IIb) at (-3,-6);
    \coordinate (Vr) at (6,3);
    \coordinate (Vt) at (3,6);
    \coordinate (Vl) at (0,3);
    \coordinate (VIt) at (-3,6);
    \coordinate (VIl) at (-6,3);

    %drawing the diagram
    \draw -- (Il) -- (Ib) --  (Ir) --  (It) -- node[midway, below, sloped] {$r_H$} (Il) -- cycle;

    \draw -- (IIl) -- (IIb) -- (Il) -- (IIt) -- (IIl) -- cycle;

    \draw -- (Vl) -- (It) -- (Vr) -- (Vt) -- (Vl) -- cycle;

    \draw -- (VIl) -- (IIt) -- (Vl) -- (VIt) -- (VIl) -- cycle;

    \node at (4.8,-4.8) {$\mathfrak{I}^-$};

    \node at (4.8,-1.1) {$\mathfrak{I}^+$};

    %singularities
    \draw[decorate,decoration=zigzag] (IIb) -- (Ib);

    \draw[decorate,decoration=zigzag] (IIt) -- (It);

    \draw[decorate,decoration=zigzag] (VIt) -- (Vt);

    %quantum region
    \tikzset{declare function={
        kruskal(\x,\c)  = {\fpeval{asin( \c*sin(2*\x) )*2/pi}};
    }}

    \draw[mydarkpurple,line width=0.4,samples=20,smooth,variable=\x,domain=-3:3] %middle bounce
        plot(\x,{-3*kruskal((\x+3)*pi/12,0.5)})
        plot(\x,{ 3*kruskal((\x+3)*pi/12,0.5)});

    \fill[fill=mylightpurple, opacity=0.7]
        plot[smooth,domain=-3:3,variable=\x] 
        (\x,{-3*kruskal((\x+3)*pi/12,0.5)})
    --
        plot[smooth,domain=-3:3,variable=\x] 
        (\x,{ 3*kruskal((\x+3)*pi/12,0.5)})
    -- cycle;

    \draw[mydarkpurple,line width=0.4,samples=20,smooth,variable=\x,domain=-3:3] %lower bounce
        plot(\x,{3*kruskal((\x+3)*pi/12,0.5)-6});

    \fill[fill=mylightpurple, opacity=0.7]
        plot[smooth,domain=-3:3,variable=\x] 
        (\x,{3*kruskal((\x+3)*pi/12,0.5)-6})
    -- (Ib) -- (IIb)
    -- cycle;

     \draw[mydarkpurple,line width=0.4,samples=20,smooth,variable=\x,domain=-3:3] %upper bounce
        plot(\x,{-3*kruskal((\x+3)*pi/12,0.5)+6});

    \fill[fill=mylightpurple, opacity=0.7]
        plot[smooth,domain=-3:3,variable=\x] 
        (\x,{-3*kruskal((\x+3)*pi/12,0.5)+6})
    -- (Vt) -- (VIt)
    -- cycle;

    \node[mydarkpurple] at (0,-1.3) {$r_{min}$};
\end{tikzpicture}
\caption{Conformal diagram to illustrate the causal structure of our Cyclic Kruskal Universe. The purple region is ``cut out" and the interior region of the black hole is glued to the interior region of the white hole along the hypersurface with coordinate $r_{\rm min}$. The exterior regions are disconnected. This chain of black and white holes can continue infinitely far above and below.}
\label{fig:conformaldiagram}
\end{figure}

Following this discussion, we arrive at a causal spacetime diagram of the form shown in Fig.~\ref{fig:conformaldiagram}. We assume that $r_{\rm min} < r_H$. The Kruskal--Szekeres coordinates only cover the usual two exterior spacetime regions and a single black-hole and white-hole interior. However, using the $(T,z)$-coordinates we can glue two Kruskal extensions together at the minimum radius hypersurface of the black/white hole respectively. This gluing produces an infinite cycle of black and white holes.

\subsection{Extending to $\Lambda_c \neq 0$}

In the previous analysis we have restricted ourselves to the case $\Lambda_c=0$, a quantum state peaked around zero cosmological constant, so that at the semiclassical level there is no cosmological constant parameter in the metric. One might of course also be interested in the more general case in which a $\Lambda$ is present.

Starting from the metric (\ref{equ:quantumss}) with (\ref{eq:etaxiexpv}) and now leaving $\Lambda_c \neq 0$ general, we can again pass to the coordinate $r(T)=\langle\hat\eta(T)\rangle$, where $\langle\hat\eta(T)\rangle$ is independent of $\Lambda_c$. We then obtain the metric form
\be
{\rm d}s^2 = - F(r)\,{\rm d}z^2 +\frac{{\rm d}r^2k_c^2\,T'(r)^2}{r^4 F(r)} + r^2\,{\rm d}\Omega^2
\label{eq:Lambdametric}
\ee
with
\begin{align}
F(r) &= 1-\frac{r_H}{r}-\frac{\Lambda_c k_c}{\sqrt{\pi}\sigma_\Lambda r}e^{-\sigma_\Lambda^2 T(r)^2}\nonumber
\\ & \quad - \frac{\Lambda_c k_c}{r} T(r) \, {\rm erf}(\sigma_\Lambda T(r)) \,.
\label{eq:F(r)}
\end{align}
The asymptotic expansion of the error function at large $|T|$ gives
\be
T \, {\rm erf}(\sigma_\Lambda T)\sim |T|-\frac{e^{-\sigma_\Lambda^2 T^2}}{\sqrt{\pi}\sigma_\Lambda}+O\left(\frac{e^{-\sigma_\Lambda^2 T^2}}{T^2}\right)
\ee
and we see that the leading correction at large $|T|$ actually cancels the last term in the first line of Eq.~(\ref{eq:F(r)}). All higher corrections are then $O\left(\frac{e^{-\sigma_\Lambda^2 T(r)^2}}{T(r)^2}\right),$ and in terms of $r$ these are at least suppressed as $O(e^{-\alpha \frac{r^6}{r_{{\rm min}}^6}}\cdot \frac{r_{{\rm min}}^6}{r^6})$ for some constant $\alpha$; we had neglected such highly suppressed terms in the earlier analysis with $\Lambda_c=0$.

The dominant corrections then come from the expansion of $|T(r)|$ at large $r$ given in Eq.~(\ref{equ:Tlarger}), and we have
\begin{align}
F(r) \sim 1-\frac{r_H}{r}-\frac{\Lambda_c r^2}{3}\left(1+\frac{\pi^3}{6\Gamma\left(\frac{2}{3}\right)^6}\frac{r_{{\rm min}}^6}{r^6}+\ldots\right)\,. \nonumber
\end{align}
Again, we see that the corrections involving $\Lambda$-dependent terms are exceedingly small when $r\gg r_{{\rm min}}$. For $\Lambda_c>0$ there is still a cosmological horizon at large $r$, whose location is very slightly shifted. The global structure of these solutions at large $r$ is again identical to those of the classical solutions, with the corrections again becoming relevant only inside the black (white) hole horizon if we assume that $r_H\gg r_{{\rm min}}$. It would be interesting to study these solutions in more detail in future work.

\section{Energy conditions}
\label{sec:energy}

Classical energy conditions (strong, weak, null, etc.), which enforce pointwise positivity of the stress-energy tensor, are known to be easily violated when quantum matter is included \cite{Epstein:1965zza}. In the context of quantum field theory (QFT) on curved spacetime, one can consider weaker energy conditions that quantum fields could satisfy (for a recent review, see Ref.~\cite{Kontou:2020bta}). It has been suggested that the energy condition that has the best chance of being satisfied in a general QFT is the achronal averaged null energy condition (AANEC) \cite{Kontou:2020bta}. The AANEC is defined by averaging the pointwise null energy condition along an achronal inextendable null geodesic. Achronal in this context means that no two points on the geodesic can be connected by a timelike curve. Put into equations, the condition is 
\begin{align}
    \int_{-\infty}^{\infty} \td\lambda \; G_{\mu\nu}(\gamma(\lambda))\dot{\gamma}^\mu(\lambda)\dot{\gamma}^\nu(\lambda) \geq 0\,,
\end{align}
where $\lambda$ is the affine parameter along a null geodesic $\gamma$. We have formulated the condition in terms of the Einstein tensor $G_{\mu\nu}$, since we focus on the geometry and we do not assume that the metric satisfies any form of the Einstein equations. In the QFT literature one usually uses the expectation value of the stress-energy tensor instead.

There are known situations in which the AANEC is violated, however, these include Planck length distances on which QFT on curved spacetime is thought to no longer be reliable \cite{Urban:2009yt}. In specific scenarios the validity of the AANEC has been proven, e.g., for free scalar fields \cite{Kontou:2015yha} or on Minkowski spacetime \cite{Flanagan:1996gw}.

In our case, a violation of the AANEC would strongly suggest that the quantum-corrected Schwarzschild spacetime cannot be interpreted as a classical black-hole background with quantum matter placed upon it.

We consider an achronal null geodesic in the $(r,z)$-plane. As in the derivation of Eq.~(\ref{equ:simpu}), we assume that $\theta=\frac{\pi}{2}$ at all times; we also consider a purely radial geodesic with no angular momentum, so $L=U^\varphi=0$. For an initially ingoing geodesic, we then obtain
\begin{align}
  U^z & = \frac{E}{1-\frac{r_H}{r}}\,,
  \\ U^r &= \frac{r^2}{k_c T'(r)}E
\end{align}
with $E>0$.

This geodesic describes a particle coming from infinite $r$ and travelling only in the $(r,z)$-plane, into the black hole, through the white hole, and back to $r=\infty$. The AANEC becomes
\begin{align}
    0\le &\int_{-\infty}^{\infty}\td \lambda\;G_{\mu\nu}(\gamma(\lambda))\dot{\gamma}^\mu(\lambda)\dot{\gamma}^\nu(\lambda)\nonumber\\
    =& \int_{\infty}^{r_{\rm min}}\td r\;\frac{1}{(U^r)_-}[G_{rr}(U^r)^2+G_{zz}(U^z)^2]\,\nonumber\\
    &+\int_{r_{\rm min}}^{\infty}\td r\;\frac{1}{(U^r)_+}[G_{rr}(U^r)^2+G_{zz}(U^z)^2]\,\nonumber\\
    =&\, \frac{4E}{k_c}\int_{r_{\rm min}}^{\infty}\td r\;\frac{r|T''(r)|-2|T'(r)|}{T'(r)^2}\,.
\label{eq:aanec}
\end{align}
By restricting $\lambda$ to the positive or negative half-line, we are able to reparametrise the corresponding half of the geodesic with the $r$ coordinate. However we have to be careful, because we get a factor of $1/U^r$ which depends on $T'(r)$. Depending on whether we are in the black-hole or white-hole part of the spacetime, we get different results for the same value of $r$, since $T(r)$ switches sign:  $(U^r)_+ = -(U^r)_-$. It follows that both halves give the same contribution to the integral, as $G_{\mu\nu}$ and $U^\mu$ are independent of $z$.

One can see that the last line of Eq.~(\ref{eq:aanec}) is independent of $\beta$. For $k=1$ and $\sigma_{\Lambda}=5$ this expression evaluates to $\approx -5.6E$, thus the AANEC is violated in our spacetime, which makes our results unlikely to be reproducible by any theory of quantum fields on a black-hole spacetime. Instead, one needs to invoke quantum aspects of the spacetime geometry itself, as in our minisuperspace approximation to quantum gravity.

The question of whether singularity resolution can be achieved without violating some generalised energy condition has been extensively discussed in the literature \cite{PhysRevD.17.2521,Borde:1987qr,PhysRevD.37.546,Fewster:2010gm}. In particular, several authors have established singularity theorems under averaged or otherwise weakened energy conditions. Given these findings, it is perhaps not unexpected that our singularity-free spacetime violates the AANEC.

\section{Conclusion}
\label{sec:conc}

Imposing unitarity with respect to unimodular time leads to a quantum minisuperspace model for the Schwarzschild spacetime in which the classical singularity is resolved \cite{Gielen:2025ovv}. Evaluating semiclassical expectation values for the metric components leads to the definition of a quantum-corrected Schwarzschild metric, which we have analysed in more detail in this work. In particular, by rewriting the metric in terms of the usual Schwarzschild coordinates we could see that the quantum modifications can be seen as stemming from the introduction of a single new length scale $r_{\rm min}$ in addition to the length scale of the horizon $r_H$ already present in the classical theory. The minimum radius $r_{\rm min}$ depends on details of the semiclassical state and dictates when the black-to-white-hole transition takes place. It is a free parameter of our model and not connected to, e.g., Planck-scale discreteness effects seen or assumed in other approaches.
The emergence of a minimal radius is also seen in models inspired by LQG, such as Ref.~\cite{Ashtekar:2018lag,Alonso-Bardaji:2022ear,Zhang:2024khj}. The causal structure of the quantum-corrected spacetime studied in Ref.~\cite{Ashtekar:2018lag,Alonso-Bardaji:2022ear,Zhang:2024khj} is notably similar to the one we have studied here: the black and white hole interiors are connected via a transition surface of minimal radius and the external regions of the black and white hole are disconnected. This causal structure was also seen in Ref.~\cite{Lewandowski:2022zce}, although the detailed modifications to the metric look different; for instance, in that model both $g_{rr}$ and $g_{zz}$ are modified. In this sense, our work shows how possible quantum gravity modifications to the classical Schwarzschild spacetime, motivated from rather different arguments within the landscape of ideas about quantum gravity, can lead to qualitatively similar effects. One might say that our scenario for the ultimate fate of a black hole is rather conservative, given that corrections only come in very close to the minimal radius where the transition to a white hole occurs, and are very strongly suppressed away from it. Our model leads to an extension of the Kruskal spacetime into the past and future, in which there is an infinite sequence of black-hole to white-hole transitions connecting disconnected, asymptotically flat universes. Just as in the usual Kruskal spacetime, this particularly simple and symmetric structure, in which the exterior spacetime is static, is not expected to represent astrophysically relevant black or white holes.

More realistic models for a quantum black (white) hole should include both the process of black hole formation via gravitational collapse and the evaporation of a black hole via Hawking radiation. Both additions would of course change the global causal structure substantially. Regarding Hawking radiation, we calculated the Hawking temperature and saw that corrections to the classical results are very small as long as $r_H \gg r_{\rm min}$. During evaporation the mass of the black hole, and thus the horizon, shrinks. What happens to the minimum radius? Below Eq.~(\ref{eq:rminM}) we argued that the minimum radius would shrink but only as $\sim M^{1/6}$. This is because $\beta$ is fixed by our quantum theory and we might also assume that the variance of $\Lambda$ stays constant. In quantum mechanics the variance of Gaussian wave packets grows over time, so it could also be that $r_{\rm min}$ decreases faster than our first approximation. In order for $r_{\rm min}$ to shrink as fast as $r_H$, $\sigma_{\Lambda}$ would have to go as $\sim M^{-5/2}$. If $r_{\rm min}$ shrinks more slowly, after some time it will be of the same order as $r_H$. In that case we expect our approximations to break down and one would need a full quantum-gravitational treatment of the problem. We could speculate, however, that the minimum radius stops the black hole from further evaporation and leads to a remnant. From the viewpoint of an external observer, Hawking evaporation implies that the black hole has a finite lifetime. Due to unitarity, however, we are forced to extend spacetime beyond the resulting naked singularity that semiclassical physics might predict. Drawing inspiration from what we found for the model discussed in this paper, it seems likely that this is done by gluing a time reversal of the evaporating black hole to the spacetime. This would again result in a black-to-white-hole transition, but with a single exterior region, again similar to what is seen in other quantum-gravity inspired models \cite{Belfaqih:2024vfk,Ashtekar:2005cj,Han:2023wxg}.

Regarding the formation of a black hole via gravitational collapse, one possible interesting comparison would be with the model of Ref.~\cite{Kelly:2020lec}, which includes LQG-inspired corrections into the Oppenheimer--Snyder collapse model. By analogy with the model discussed in this paper, we expect unitarity in unimodular time to prevent a collapsing star from ending in a singularity. The causal structure found in Ref.~\cite{Kelly:2020lec} differs substantially from ours, since the exterior regions of the collapsing and expanding star are connected. It would be intriguing to see if this also happens in our model if we introduce dynamics instead of having a static exterior. 

When it comes to energy conditions, the authors of \cite{Alonso-Bardaji:2022ear,Ashtekar2023} discuss classical energy conditions, which are violated in their spacetime, but do not consider more general or averaged energy conditions. As far as we can tell, averaged energy conditions for LQG-inspired regular black hole models have not been discussed in the literature. However, it was shown that loop quantum cosmological spacetimes can violate the averaged null energy condition \cite{Li:2008sw}, and the AANEC was investigated in a black hole-white hole model related to LQG-inspired models \cite{Bardeen:2018frm,Bardeen:2020lko}. We have demonstrated that our model violates the achronal averaged null energy condition, which is generally believed to hold for physically reasonable quantum field theories \cite{Kontou:2020bta}. This suggests the model captures quantum geometric features that cannot be accessed via semiclassical approaches. This is not surprising, as the singularity is resolved by boundary conditions associated to unitarity in the quantum theory.

\

{\em Acknowledgements} -- The work of SG is funded by the Royal Society through the University Research Fellowship
Renewal URF$\backslash$R$\backslash$221005.

\

{\em Data Availability Statement} -- The data that support the findings of this article are openly available \cite{arXivversion}.

\bibliographystyle{JHEP.bst}
\bibliography{quantummetric}

\end{document}